\documentclass[twocolumn]{aa}
\usepackage{graphicx}
\usepackage{txfonts}
\begin{document}
   \title{A companion to AB\,Pic at the planet/brown dwarf boundary\thanks{Based on
ESO observing programs 70.C-0677(B) and 072.C-0644(B) and 073.C-0469(B) at the VLT}}

   \subtitle{}

\author{
        G. Chauvin\inst{1}\and
        A.-M. Lagrange\inst{2}\and
        B. Zuckerman\inst{3}\and
        C. Dumas\inst{1}\and
        D. Mouillet\inst{4}\and
	I. Song\inst{3}\and
        J.-L. Beuzit\inst{2}\and
        P. Lowrance\inst{5}\and
	M.\,S. Bessell\inst{6}
}

 \offprints{Ga\"el Chauvin \email{gchauvin@eso.org}}
\institute{
$^{1}$European Southern Observatory, Casilla 19001, Santiago 19, Chile\\
$^{2}$Laboratoire d'Astrophysique, Observatoire de Grenoble, 414, Rue de la piscine, Saint-Martin d'H\`eres, France\\
$^{3}$Department of Physics \& Astronomy and Center for Astrobiology, University of California, Los Angeles, Box 951562, CA 90095, USA\\
$^{4}$Laboratoire d'Astrophysique, Observatoire Midi-Pyr\'en\'ees, Tarbes, France\\
$^{5}$Spitzer Science Center, Infrared Processing and Analysis Center, MS 220-6, Pasadena, CA 91125, USA\\
$^{6}$ Research School of astronomy and Astrophysics Institue of Advance Studies, Australian National University, Cotter Road, Weston Creek, Canberra, ACT 2611, Australia
}

   \date{Received September 15, 1996; accepted March 16, 1997}

   \abstract{We report deep imaging observations of the young, nearby star AB\,Pic,  a member of the large Tucana-Horologium association. We have detected a faint, red source $5.5~\!"$  South of the star with JHK colors compatible with that of a young substellar L dwarf. Follow-up observations at two additional epochs confirm, with a confidence level of 4.7$\sigma$, that the faint red object is a companion to AB\,Pic rather than it being a stationary background object. A low resolution K-band spectrum indicates an early-L spectral type for the companion. Finally, evolutionary model predictions based on the JHK photometry of AB\,Pic\,b indicate a mass of 13 to 14 MJup if its age is $\sim30$~Myr. Is AB\,Pic\,b a massive planet or a minimum mass brown dwarf? }

   \maketitle
%

\section{Introduction}

In October 2000, we began a deep imaging survey of stars in young, nearby
southern associations to search for companion brown dwarfs and giant
planets.  First, we used the ADONIS/SHARPII adaptive optics (AO)
instrument at the ESO/3.6 m telescope (Chauvin et al. 2003). Since
November 2002 we have pursued the survey with the VLT/NACO instrument. We
mainly focussed our search on the Tucana-Horologium (Torres et
al. 2000, Zuckerman \& Webb 2000) and TW Hydrae (Kastner et al. 1997) associations, as
well as the $\beta$ Pictoris (Zuckerman et al. 2001) and
AB Doradus (Zuckerman et al. 2004) co-moving groups, to explore the circumstellar environment within semimajor axes between tens and hundreds of AU with detection limits down to a few Jupiter masses.

This strategy resulted in the astrometric and spectroscopic confirmation
of a brown dwarf companion to GSC\,08047-00232 (Chauvin et al 2005,
Neuh\"auser \& Guenther 2004) and detection of a giant planet companion
candidate to the young brown dwarf 2M1207 (Chauvin et al. 2004; Chauvin et al. 2005, submitted). 

In March 2003, we observed the young star AB\,Pic (HIP\,30034, K2V, V=9.16,
$d=47.3_{-1.7}^{+1.8}~$pc), identified by Song et al. (2003) as a member
of the large Tucana-Horologium association of estimated age $\sim30$~Myr,
according to a comparison of $\rm{V}-\rm{K}$ versus $\rm{M}_{K}$ with
evolutionary tracks. At 5.5~$\!''$ South from AB\,Pic, we detected
(Fig.~1) a faint and red object with a near-IR color compatible with
that observed for cool L dwarfs (Leggett et al. 2002). We re-observed both
objects in March and September 2004 in order to determine if they
shared common proper motions. In December 2004, acquisition of a near-IR
spectrum enabled us to determine the spectral type of AB\,Pic\,b and to
confirm its substellar status.

%

\section{Observations}

\begin{table*}[t]
\caption{Observing Log}
\label{tab:setup}
\centering
\begin{tabular}{llllllllll}     
\hline\hline
Name        & UT Date       &   Tot. Exp. time  &  Filter  & Camera  &  Mode            &Strehl   &Seeing   &Airmass & Remarks\\
        &       &     &    &   &              &(\%)   &(arcsec)    & & \\
\noalign{\smallskip}\hline\noalign{\smallskip}
\multicolumn{10}{c}{Classical imaging and coronagraphy}\\
\noalign{\smallskip}\hline\noalign{\smallskip}
AB\,Pic\,A & 17/03/2003  & 60$\times$2s        & NB1.24  & S13      & classical       & 12      & 0.75     & 1.32    & science   \\
              & 17/03/2003  & 120$\times$0.5s     & NB1.75  & S13      & classical       & 25      & 0.80     & 1.22    & science      \\
              & 17/03/2003  & 100$\times$0.35s    & NB2.17  & S27      & classical       & 32      & 0.80     & 1.31    & science     \\
AB\,Pic\,b & 17/03/2003  & 18$\times$30s       & J       & S13      & coronagraphy    & 12      & 0.75     & 1.27    & science   \\
              & 17/03/2003  & 5$\times$30s        & H       & S13      & coronagraphy    & 25      & 0.80     & 1.28    & science   \\
              & 17/03/2003  & 3$\times$20s        & K$_{s}$ & S27      & coronagraphy    & 32      & 0.80     & 1.30    &science   \\
$\theta$ Ori 1\,C& 16/03/2003  & 12$\times$10s    & NB1.75  & S13      & coronagraphy    & 37      & 1.00     & 1.12    &    astrometric std     \\
  \noalign{\smallskip}\noalign{\smallskip}
AB\,Pic\,A & 05/03/2004  & 10$\times$1s        & NB1.75  & S13      & classical       & 20      & 1.00    & 1.20    & science      \\
AB\,Pic\,b & 05/03/2004  & 4$\times$30s        & H       & S13      & coronagraphy    & 20      & 1.10    & 1.20    & science      \\
$\theta$ Ori 1\,C & 05/03/2004  & 10$\times$12s   & H       & S13      & coronagraphy    & 38      & 1.10    & 1.15    & astrometric std       \\
\noalign{\smallskip}\noalign{\smallskip}
AB\,Pic\,A & 26/09/2004  & 2$\times$5s         & H+ND    & S13      & classical       & 27      & 0.74   & 1.28    & science      \\
AB\,Pic\,b & 26/09/2004  & 3$\times$20s        & H       & S13      & coronagraphy    & 27      & 0.75   & 1.28    & science      \\
$\theta$ Ori 1\,C & 26/09/2004  & 4$\times$0.8s   & H       & S13      & coronagraphy    & 24      & 0.85     & 1.15  &   astrometric std       \\
\noalign{\smallskip}\hline\noalign{\smallskip}
\multicolumn{10}{c}{Spectroscopy}\\
\noalign{\smallskip}\hline\noalign{\smallskip}
AB\,Pic\,b & 03/12/2004  & 16$\times$300s     & SHK      & S54      & $\rm{R}_{\lambda}=550$    & na      & 0.87   & 1.26       & science      \\
HIP\,33632   & 03/12/2004  & 16$\times$300s     & SHK      & S54      & $\rm{R}_{\lambda}=550$    & na      & 0.70    & 1.08       & telluric std       \\
\hline
\end{tabular}
\end{table*}

\begin{figure}[!b]
\centering
\includegraphics[width=6.3cm]{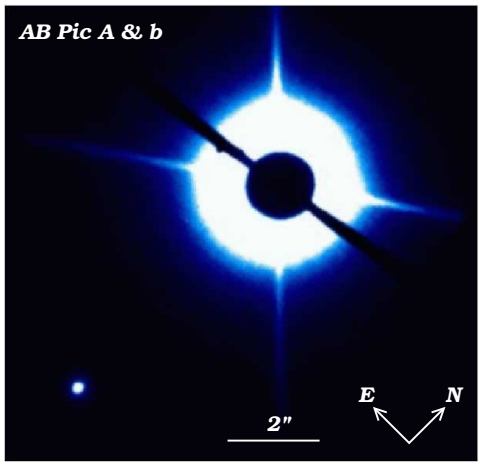}
\caption{K$_{s}$-band coronagraphic image of AB\,Pic\,A and b acquired on 17
March 2003 with an  occulting mask of diameter 1.4~$\!''$.}
\label{Fig:image}
\end{figure} 
AB\,Pic\,A and b were imaged with NACO using classical imaging and
coronagraphy (Table 1 and Fig.~2). The JHK photometry of AB\,Pic\,A is known
from the 2MASS All-Sky Catalog of Point Sources (Cutri et al. 2003). On 17 March 2003,
we measured the JHK contrasts of AB\,Pic\,A and b, using the
deconvolution algorithm of V\'eran \& Rigaut (1998), so as to determine the photometry of
AB\,Pic\,b (see Table~2). The transformation between the K$_{s}$ filter of
NACO and the K filter used by CTIO-2MASS was found to be smaller than 0.03
magnitude. On 5 March 2004 and 26 September 2004, follow-up observations of AB\,Pic\,A
and b were acquired in the H band with the S13 camera to provide the best
compromise between high angular resolution and AO correction. To calibrate the
platescale and the detector orientation, we observed at each epoch the
astrometric field of $\theta$ Ori 1 C. The orientations of true north of the
S13 camera were found on 16 March 2003, 5 March 2004 and 26 September 2004
respectively at $-0.05^o, 0.04^o, 0.20^o$ east of the vertical with an
uncertainty of $0.10^o$. The pixel scale was found to be relatively stable in
time with values of $13.21\pm0.11$\,mas, $13.24\pm0.05$\,mas and
$13.23\pm0.09$\,mas.

The NACO spectroscopic observations of AB\,Pic\,b were obtained on 3
December 2004, using the low resolution ($\rm{R}_{\lambda}=550$) grism
with the 86~mas slit, the S54 camera (54~mas/pixel) and the SHK filter covering the
entire spectral range between 1.39 and 2.52~$\mu$m. The telluric standard star
HIP\,33632 (B6V) was also observed. After substracting the sky and
dividing by a flat field using \textit{eclipse} (Devillard 1997),  the spectra
of AB\,Pic\,b and HIP\,33632 were extracted and calibrated in wavelength
with \textit{IRAF/DOSLIT}.
To calibrate the relative throughput of the
atmosphere and the instrument, we divided the extracted spectrum of
AB\,Pic\,b by the spectrum of HIP\,33632. To restore the continuum
shape, we then multiplied by a composite spectrum of a B6IV
star taken from a library of stellar spectra (Pickles 1998).

\begin{table}[b]
\caption{Photometry of AB\,Pic\,A and b}
\label{table:1}
\centering
\begin{tabular}{llll}     
\hline\hline
Component         &  J  & H    &  K \\
             &  (mag)         &  (mag)    & (mag)               \\
\hline
AB\,Pic\,A$^{a}$ & 7.58 +- 0.03 & 7.09 +- 0.03  & 6.98 +- 0.03      \\
AB\,Pic\,b$^{b}$ &16.18 +- 0.10 & 14.69 +- 0.10 & 14.14 +- 0.08     \\
\hline
\end{tabular}
\begin{list}{}{}
\item[$^{\mathrm{a}}$] from the 2MASS All-Sky Catalog of Point Sources (Cutri et al. 2003).
\item[$^{\mathrm{b}}$] from $^{\mathrm{a}}$ and NACO measurements presented in this work.
\end{list}
\end{table}

%
\section{Companionship confirmation}

\begin{figure}[t]
\centering
\includegraphics[width=9.cm]{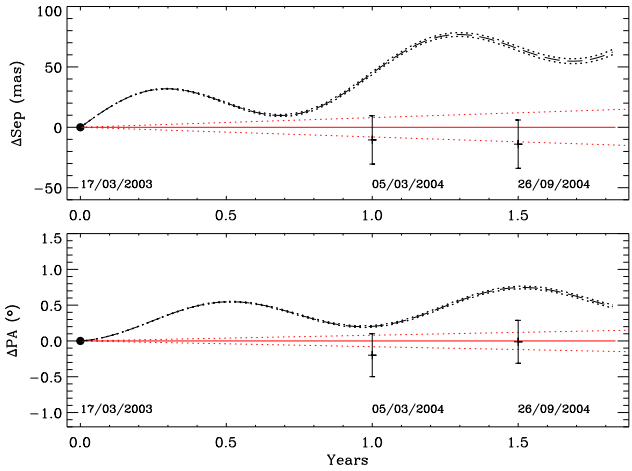}
\caption{\textbf{Top:} Differences of separation between
AB\,Pic A and b on 17 March 2003, 5 March 2004, and 26 September
2004. \textbf{Bottom:} Differences of position angle for AB\,Pic A and b.
The expected evolutions with their uncertainties for bound (\textit{solid
line}) and stationary background (\textit{dashed line}) objects are indicated.}

\label{Fig:astro}

\end{figure}

To verify that AB\,Pic\,A and b were comoving together in the sky and thus
physically bound, their relative positions were determined on 17 March 2003, 5
March 2004 and 26 September 2004 (see Table~3). We then took into account the
proper motion of AB\,Pic\,A from the Tycho catalog \cite{hog00}:
$\mu_{\alpha}=15.9\pm1.2$~mas/yr and $\mu_{\delta}=46.2\pm1.2$~mas/yr, its expected parallactic motion and
the detector calibrations at each epoch (platescale and detector orientation,
see section~2). The expected variations in separation and position angle
in the case of a bound companion and of a background stationary object are
shown in Fig.~2. The maximal orbital motion of AB\,Pic\,b from March 2003
to September 2004 is $<12$~mas. In the case of a stationary background object, important variations are expected in both separation and position angle. Between March 2003 and September 2004, differences of separation and position angle show that AB\,Pic\,b is not a stationary background object, at the 4.7\,$\sigma$ confidence level.
\begin{table}[h]
\caption{Astrometric measurements of AB\,Pic\,A and b and confidence level estimation that AB\,Pic\,b is not a stationary background object.}
\label{table:1}
\centering
\begin{tabular}{llll}     
\hline\hline                      
UT Date      &    Separation     &    Position Angle         &  Confidence\\
             &    (mas)          &  ($^o$)         & Level \\
\noalign{\smallskip}\hline\noalign{\smallskip}
 17/03/2003             & $5460\pm14$       & $175.33\pm0.18$         &-  \\
 05/03/2004             & $5450\pm16$       & $175.13\pm0.21$        &3.0\,$\sigma$  \\
 26/09/2004             & $5453\pm14$       & $175.10\pm0.20$        &4.7\,$\sigma$  \\
\hline
\end{tabular}
\end{table}
%
\section{Spectral type determination}

Based on the near-IR photometry presented in Table~2, we can derive the near-IR
colors of AB\,Pic\,b: $(J-K)=2.04$, $(J-H)=1.49$ and $(H-K)=0.55$. These
values are consistent with that observed by Knapp et al. (2004) for late-type L
dwarfs.

From spectroscopy, we decided to consider only the K-band portion of the
AB\,Pic\,b spectrum which is less subject than the H-band to chromatical
effects, mainly due to differential slit centering and/or differential AO
corrections between the science source and the telluric standard (see Goto et al. 2002; Chauvin et al. 2004). Based on
the template spectra for late-M, L and T dwarfs of Leggett et al.
(2001) and Geballe et al. (2002), we used a minimum $\chi^2$ adjustment to find
the best template spectra matching the broad water-band absorptions of the
AB\,Pic\,b spectrum. This allowed us to derive a spectral type
L$1^{+2}_{-1}$ for AB\,Pic\,b. The best adjustment is obtained with the L1
dwarf 2MASS\.J0345+2540 (see Fig.~3). In addition, various water band indices confirm this spectral type
estimation. The water band index Q of Wilking et al. (1999) is equal to 0.44
for AB\,Pic\,b which confirms a spectral type later than M9. The K1 and K2
indices of Reid et al. (2001) are respectively equal to $0.23$ and $-0.03$. The
H$_2$O 2.0$\mu$m index of Geballe et al. (2002) is equal to 1.15. They all
confirm a spectral type L0 to L3 for AB\,Pic\,b. The presence of absorption
lines of NaI~(2.205 and 2.209\,$\mu$m) and the $^{12}$CO transitions 2-0
(2.295\,$\mu$m), 3-1 (2.324\,$\mu$m), 4-2 (2.354\,$\mu$m) and 5-3
(2.385\,$\mu$m) also support the fact that AB\,Pic\,b is an early-L dwarf.

\begin{figure}[t]
\centering
\includegraphics[width=8.cm]{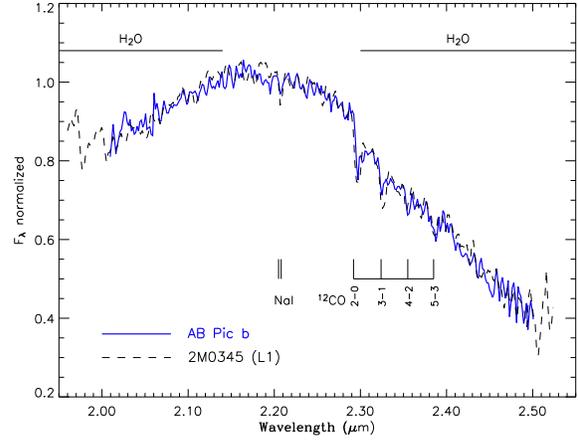}

\caption{K-band spectrum of AB\,Pic\,b acquired on 3 December 2004 with the
low resolution ($\rm{R}_{\lambda}=550$) grism of NACO, the 86~mas slit and the
S54 camera (54~mas/pixel). The best $\chi^2$ adjustment is found with the L1
dwarf 2MASS\.J0345+2540 (Geballe et al. 2002).}

\label{Fig:image}
\end{figure}

%

\section{Temperature and mass determination}

As an L$1^{+2}_{-1}$ dwarf companion to a
30~Myr old star, AB\,Pic\,b is unambiguously a substellar object. To
investigate its physical properties (mass, temperature and luminosity),
evolutionary model predictions can be compared to the absolute photometry of
AB\,Pic\,b.  However, as described by Baraffe et al. (2002), model
predictions must be considered carefully as they are still uncertain at early
phases of evolution ($\le100$~Myr; see also Mohanty et al. 2004 and Close et al. 2005). 


We then considered the most commonly used models of Burrows et al. (1997), Chabrier et al. (2000) and Baraffe et al. (2002) to describe the
atmospheric properties of young brown dwarfs and giant planets.
Table 4 displays predictions of DUSTY models of Baraffe, of Burrows et al (1997) and B. Macintosh \& M. Marley
(personal communication).  As may be seen, for a 30 Myr old object, the two
calculations agree rather well, with one a bit redder in J-K than AB\,Pic\,b and the other a bit bluer.

\begin{table}[t]
\label{table:1}      
\centering          
\caption{Absolute magnitudes of AB\,Pic\,b compared to evolutionary model predictions (Burrows et al 1997; Baraffe et al 2002; B. Macintosh \& T. Barman, private communication).}
\begin{tabular}{lllllll}     
\hline\hline                      
\small{Model}		&Age	&Mass	&T$_{\rm{eff}}$	&M$_{J}$	&M$_{H}$	&M$_{K}$   \\
		&(Myr)	&(M$_{\rm{Jup}}$)	&(K) &(mag)  & (mag)       &   (mag)   \\
\noalign{\smallskip}\hline\noalign{\smallskip} 
AB\,Pic\,b	&30	&	& 	&12.80	&11.37	&10.76\\
\noalign{\smallskip}\hline\noalign{\smallskip} 
Burrows$\,^{*}$		&30.4	&13	&1513	 &12.8	&11.9	&11.2\\
		&30.4	&14	&1856	& 11.5	&10.7	&10.1\\
\noalign{\smallskip}\hline\noalign{\smallskip} 
DUSTY$\,^{*}$	&30	&13	&1594	& 14.0	&12.2	&11.0\\
		&30	&14	&1764	& 12.8	&11.5	&10.5\\
		&20	&13	&1672	& 13.2	&11.7	&10.7\\
		&40	&14	&1701	& 13.3	&11.8	&10.7\\
\hline                  
\end{tabular}
\begin{list}{}{}
\item[$^{\mathrm{*}}$] The gravity factor log\,(\textit{g}) is respectively equal to 4.0 and 4.1 for the Burrows and DUSTY models.
\end{list}
\end{table}


\section{Gravitational collapse or core accretion?}

With a model-derived mass of about 13~M$_{\rm{Jup}}$, one can speculate whether AB\,Pic\,b is a giant planet or a brown dwarf. The \textit{International Astronomical Union} has recently adopted a definition to differentiate planetary and brown dwarf companions. The latter are objects with masses above the minimum mass for deuterium burning (13.6~M$_{\rm{Jup}}$). Based on this criterion, as one may see from Table 4, it is going to be very difficult to decide between very high mass planet or a very low mass brown dwarf, given uncertainties in evolutionary models and in the age of AB\,Pic\,b. A second criterion, defended by an important fraction of the community, would be to differentiate planet from brown dwarf according to their origins of formation. But, more than discussing the definition of AB\,Pic\,b, everyone should agree that a more meaningful question is here to know whether this object has formed by gravitational collapse, that is, in a one-step process, or by a two-step process that begins with core accretion. 

An answer is suggested by early results from imaging surveys for young planets in combination with data from the precision radial velocity technique (PRV). In the very close stellar environment, closer than  $\sim4\,$AU, PRV results indicate a mass spectrum that increases toward lower masses, from 10~M$_{\rm{Jup}}$ to less than 1~M$_{\rm{Jup}}$ (see Jorissen et al. 2001). In contrast, our VLT survey of $\sim50$ stars, which is sensitive to masses down to about 2~M$_{\rm{Jup}}$ at physical separations wider than $\sim80$~AU, has revealed only the possible planetary mass companion AB\,Pic\,b ($\sim13$ M$_{\rm{Jup}}$ at $\sim260$~AU). (The case of 2M1207\,b is not considered in the discussion as the primary is a young brown dwarf). In addition, no other sensitive near-infrared imaging survey of young stars -- with VLT/NACO (Masciadri et al. 2005), with the Keck AO system (Macintosh et al., in preparation) and with HST (Song et al.,in preparation) -- has yet reported any objects of planetary mass. Thus, this apparent absence of companions of a few Jupiter masses, suggests that wide companions like AB\,Pic\,b have formed in situ by gravitational collapse. In addition, for separations as large as $\sim260$~AU, formation by core accretion of planetesimals seems very unlikely because of inappropriate timescales to form planetesimals at such distances (Augereau et al. 2001). Gravitational instabilities within a protoplanetary disk (Papaloizou \& Terquem 2001; Rafikov 2005) or mechanisms proposed for brown dwarfs formation (Kroupa \& Bouvier 2003) appear to be more probable scenarii to explain the origin of wide companions such as AB\,Pic\,b. 

Interestingly, gravitational collapse mechanisms, which are relatively insensitive to metallicity, may also be true for very massive PRV planets with small semi-major axes ($<4\,$AU). Inspection of stellar metallicity vs planet mass data given in Santos et al (2004) and Fischer \& Valenti (2005) indicates that the well-known correlation between high stellar metallicity and the existence of planets may not be present for stars with the highest mass planets ($>7~$M$_{\rm{Jup}}$; see also Rice et al. 2003). That is, these relatively few highest mass PRV planets may have formed by gravitational collapse, while the lower mass PRV planets formed mostly or entirely via core accretion with subsequent gravitational infall. As the number of systems, detected by PRV and direct imaging, continues to grow, it should be possible to confirm or deny these tentative correlations.


%

\begin{acknowledgements}
We would like to thank the staff of the ESO, Gilles Chabrier, Isabelle Baraffe and France Allard for providing the latest update of their evolutionary models, Sandy Leggett and Tom Geballe who kindly sent us their near-infrared template spectra, and finally Sandy Leggett again for her remarks on the letter. 
\end{acknowledgements}

\end{document}